\documentstyle[aps,epsfig]{revtex}

\begin{document}

\twocolumn[
 \hsize\textwidth\columnwidth\hsize
 \csname@twocolumnfalse\endcsname

\draft
\title{Efficiency of Mesoscopic Detectors}
\author{S. Pilgram$^1$ and M. B\"uttiker$^1$}

\address{$^1$Dept. Phys. Th\'eorique, Universit\'e de Gen\`eve,
24, quai Ernest-Ansermet, 1211 Gen\`eve 4, Switzerland}

\maketitle

\begin{abstract}
We consider a mesoscopic measuring device whose conductance
is sensitive to the state of a two-level system.
The detector is described with the help of its scattering matrix.
Its elements can be used to
calculate the relaxation and decoherence
time of the system, and determine the characteristic
time for a reliable measurement.
We derive conditions needed for an efficient ratio of decoherence
and measurement time. To illustrate the
theory we discuss the distribution function of the efficiency of an
ensemble of open chaotic cavities.                                        
\end{abstract}

]


Mesoscopic physics is evolving toward a stage 
where the understanding of the measurement process 
becomes of increasing importance. Of interest are 
detectors which allow a fast determination of the state 
of the system but at the same time 
leave the coherence of the measured system as unaffected
as possible. These are conflicting requirements: For instance 
a tunnel contact is an efficient  but slow detector. 
Therefore, the question arises whether it is possible 
to develop detectors which are both fast and efficient. 
To answer this question we investigate mesoscopic multichannel 
conductors and analyze their speed and efficiency. 
The efficiency of a detector is determined by 
the ratio of the measurement time and the decoherence time
of the measured system. 

The effect of a detector on a phase coherent 
mesoscopic system has been elegantly demonstrated 
in recent experiments \cite{Buks1,Sprinzak1,Field1}.
Theoretical discussions addressed
different aspects of weak measurement in mesoscopic systems:
the relation to scattering theory \cite{Buks1,Stodolsky1,MBMartin1}
and screening \cite{MBMartin1},
the measurement time and interactions \cite{Aleiner1} 
and the relation between
detector noise and decoherence rate \cite{Levinson1,Levit1}. 
The time evolution of system and detector has been
studied using a master equation approach \cite{Gurvitz1,Schoen1}. 
Refined calculations consider the conditional evolution of
the system depending on the outcome of the measurement 
\cite{Korotkov1,Milburn1}.
Tunnel contacts and single electron transistors
have been identified as candidates for efficient
measurement devices \cite{Averin2}.

Both the measurement time and the decoherence rate depend 
on the scattering matrix of the conductor (detector). 
Consequently both of these quantities depend on the 
sample specific geometry and impurity distribution of the detector. 
It is therefore necessary to investigate the distribution 
of the quantities of interest (measurement time, decoherence rate 
and efficiency) of ensembles of macroscopically identical detectors. 
Here we focus on ballistic detectors for which ensemble 
members differ only in their geometry. 

\begin{figure}[htb]
\begin{center}
\leavevmode


\psfig{file=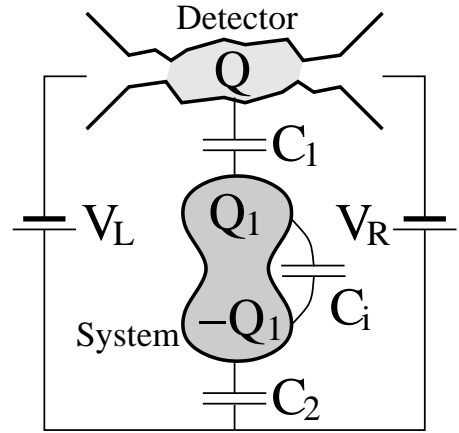,width=6cm}

\vspace{5mm}

\caption{A mesoscopic detector is capacitively coupled
to one side of a double dot. 
}
\label{Sketch1}
\end{center}
\end{figure}

The model we consider is shown in Fig. \ref{Sketch1}.  
It consists of a
double dot (DD) that plays the role of 
the system. It is an effective two-level system:
The topmost electron in the DD can either occupy the upper or the lower
dot. The detector is a mesoscopic two-terminal conductor (MC):
Its conductance is sensitive to the charge on
the upper dot. The coupling between system and detector is described
by a set of capacitances $C_1,C_2,C_i$ that link the charge $Q$ on the MC
to the charges on the dots $Q_1$ and $-Q_1$ (we abbreviate
$C^{-1} = C_1^{-1} + C_2^{-1} + C_i^{-1}$).

The interaction of DD and MC can be investigated from two different
viewpoints. From the system side we are interested in the question
of how fast a pure state prepared in the two-level system decays into
a statistical mixture. We distinguish the thermal {\it relaxation} 
to an equilibrium distribution (described by a rate $\Gamma_{rel}$) and the
often much faster {\it decoherence} of 
superpositions of states in the upper and lower dot
(described by a rate $\Gamma_{dec}$). This decoherence
depends on the temperature $kT$ as well as on the potential 
difference $eV$ applied
at the MC. From the detector side we may ask how long it takes to
{\it measure} the state
of the two-level system (described by a rate $\Gamma_m$). 
The decoherence rate at zero temperature $\Gamma_v$
is intimately related to the measurement rate $\Gamma_m$ and satisfies
the inequality $\Gamma_v \ge \Gamma_m$ \cite{Schoen1,Averin1}.

We are interested in the conditions under which 
a MC turns out to be an efficient detector, i.e. fulfills 
$\Gamma_v \gtrsim \Gamma_m$.
In order to be able to describe a
wide class of detectors we represent the MC by a scattering matrix
$s_{\alpha\beta}$ that connects in- and outgoing states ($\alpha,\beta$
label left and right reservoir). This enables us to treat
multi-channel MCs with arbitrary transmission probabilities $T_n$
and to include screening effects between different channels.
On the other hand a minimum effort is put into the description
of the coupling between system and detector. We use a standard master
equation (Bloch-Redfield approach \cite{Carmichael1}) in lowest order
perturbation theory to study the evolution of the reduced density
matrix of the DD. On this level of approximation the dynamics of the
DD is influenced by the charge fluctuation spectrum $S_{QQ}$
of the MC. A crucial role is therefore attributed to the
Wigner-Smith time delay matrix ($\beta\gamma$ label the reservoirs)
\begin{equation}
\label{Density Elements}
N_{\beta\gamma} = \frac{1}{2\pi i} \sum_{\alpha}
s_{\beta\alpha}^{\dagger} \frac{ds_{\gamma\alpha}}{dE}.
\end{equation}
that characterizes fully the low-frequency charge
fluctuations \cite{Pedersen1}.
We introduce the following four constants ($e$ denotes the
electron charge)
\begin{equation}
\label{Four Parameters}
\begin{array}{cc}
D = e^2 \mbox{Tr} N, &
C_{\mu}^{-1} = C^{-1} + D^{-1},\\
\quad\\
R_q = \frac{1}{2}\frac{\left(\mbox{Tr} N^2\right)}{\left(\mbox{Tr} N\right)^2}, &
R_v = \frac{\left(\mbox{Tr} N_{12}N_{21}\right)}{\left(\mbox{Tr} N\right)^2}.\\
\end{array}
\end{equation}
These constants have been applied in many different contexts such as
ac-transport and noise \cite{Buettiker2,Buettiker1}. $D$ corresponds
to the density of states at Fermi energy in the scattering region,
$C_{\mu}$ is an effective electrochemical capacitance that
characterizes the strength of interaction, $R_q$ expresses the
equilibrium contribution to the charge fluctuation spectrum $S_{QQ}$,
and $R_v$ the non-equilibrium contribution.

The two-level system
is conventionally represented by the Hamiltonian 
$\hat{H}_{DD} = \frac{\epsilon}{2}\hat{\sigma}_z 
 + \frac{\Delta}{2} \hat{\sigma}_x$ where
$\hat{\sigma}_i$ denote Pauli matrices.
The energy difference between upper and lower dot is $\epsilon$ and $\Delta$
accounts for tunneling between the dots. The full level splitting is
thus $\Omega=\sqrt{\epsilon^2+\Delta^2}$.

For the relaxation and decoherence rate in the DD we find the following
expressions:
\begin{equation}
\label{Central Result Relaxation}
\begin{array}{rl}
\Gamma_{rel} = & 2\pi \frac{\Delta^2}{\Omega^2}
\left(\frac{C_{\mu}}{C_i}\right)^2
R_q \frac{\Omega}{2}\coth \frac{\Omega}{2kT},\\
\end{array}
\end{equation}
\begin{equation}
\label{Central Result Decoherence}
\begin{array}{rl}
\Gamma_{dec} = & 
 2\pi \frac{\epsilon^2}{\Omega^2}
\left(\frac{C_{\mu}}{C_i}\right)^2
\left(R_q kT + R_v e|V|\right) + \Gamma_{rel}/2.\\
\end{array}
\end{equation}
Eq. (\ref{Central Result Relaxation},\ref{Central Result Decoherence})
are the central result of this paper. It has formally
the same appearance as the rates given in \cite{Schoen1}. Its big
virtue lies in the fact that the structure of the detector
is condensed into the four parameters given in (\ref{Four Parameters}).
An analysis of its properties reduces hence to a discussion of
a few parameters. We postpone this discussion
and explain first the derivation of Eqs. 
(\ref{Central Result Relaxation},\ref{Central Result Decoherence})
in order to clarify the approximations made.

The Coulomb energy of our model can be found by circuit analysis
\begin{equation}
\hat{H}_C = \frac{(\hat{Q}_1-\bar{Q}_0)^2}{2C_i} + \frac{\hat{Q}_1\hat{Q}}{C_i}
          + \frac{\hat{Q}^2}{2C}.
\end{equation}
Its first term contributes to the level splitting
of the DD ($\bar{Q}_0$ is a background charge depending on the
applied voltage $(V_L+V_R)/2$). The charging energy $e^2/2C_i$
must be large compared to $kT,e|V|$ to allow us to
consider only two levels of the DD. The second term
$\hat{Q}_1\hat{Q}/C_i$ couples system and detector. In the
derivation of the master equation for the reduced density
matrix of the system we assume weak coupling and treat this term
perturbatively. 
We apply a Markov approximation which is strictly speaking
only valid at long time scales (compared to the correlation time
of the detector)
and therefore get pure exponential relaxation and decoherence.
The third term influences the
fluctuation spectrum of the charge operator $\hat{Q}$.
In contrast to earlier work (with the exception of Ref. 
\cite{MBMartin1}) we do not
completely disregard this term, but include it 
on the level of RPA. This Gaussian approximation
restricts us to geometries, where Coulomb blockade
effects are weak.

We will now discuss the meaning of the parameters
given in Eq. (\ref{Four Parameters})
which describe the relation between detector geometry and
relaxation or decoherence on the DD.

The parameter $R_q$ lies
always in the range $1/2 > R_q > 1/2N$ where $N$ is the
dimension of the scattering matrix. This observation
indicates already that the relaxation and decoherence rates
$\Gamma_{rel},\Gamma_{dec}$ do not simply scale with
the number of channels through the system.
It is important to note that the multichannel result 
for the relaxation and decoherence rates cannot be
obtained as a sum of rates
due to each channel. For a large number $N$ of open
channels $R_q$
behaves as $\ln(N)/N$ whereas the electrochemical
capacitance $C_{\mu}\rightarrow C$ tends to a constant.
The constant $R_v$ decreases even stronger than $R_q$ like $1/N$.
We find therefore the somewhat surprising result
that relaxation and decoherence decrease in the large
channel limit. This result is a consequence of
screening in the MC which reduces the charge fluctuations
with increasing channel number $N$.

It is interesting to note that the charge response 
$D \propto \sum \tau_n$ can be expressed entirely by the
dwell times $\tau_n$ which
are eigenvalues of the matrix (\ref{Density Elements}).
This is also the case for the thermal
relaxation parameter $R_q \propto
\sum \tau_n^2 / (\sum \tau_n)^2$. For the thermal
charge fluctuations it is unimportant whether a
scattering state is connected to the left or the
right reservoir. This is reflected by the fact that
the trace in the definition of $R_q$ (see (\ref{Four Parameters}))
has to be taken over the entire matrix $N$.

On the contrary, $R_v$ does not show this symmetry.
An applied voltage distinguishes the two reservoirs from one 
another. Thus, the trace in $R_v$ (see
(\ref{Four Parameters}))
cannot be expressed by dwell times only.
To clarify the origin of $R_v$ we apply a
basis transformation in each reservoir and
divide the scattering matrix in $2\times 2$
blocks of the form
\begin{equation}
s^{(n)} = \left(
\begin{array}{cc}
-i\sqrt{R_n}e^{i(\phi_n+\phi_{A,n})} & 
\sqrt{T_n}e^{i(\phi_n-\phi_{B,n})}\\
\sqrt{T_n}e^{i(\phi_n+\phi_{B,n})} & 
-i\sqrt{R_n}e^{i(\phi_n-\phi_{A,n})}\\
\end{array}
\right).
\end{equation}
Each block is defined by its transmission
probability $T_n=1-R_n$ and three scattering
phases $\phi_n$,$\phi_{A,n}$,$\phi_{B,n}$.
Using the definition of $R_v$ (Eq. (\ref{Four Parameters})
we arrive at \cite{Pedersen1}
\begin{equation}
\label{Polar Decomposition}
R_v =
\frac{\sum_n \left(
\frac{1}{4 T_n R_n} \left(\frac{dT_n}{dE}\right)^2
+ T_n R_n \left(\frac{d\phi_{A,n}}{dE}+\frac{d\phi_{B,n}}{dE}\right)^2
\right)}
{\left(\sum_n \frac{d\phi_n}{dE}\right)^2}.
\end{equation} 
Eq. (\ref{Polar Decomposition}) can be connected to earlier 
results \cite{Buks1,Aleiner1,Levinson1} 
in the infinite capacitance limit where $C^{2}_{\mu} R_v$ 
in Eq. (4) can be replaced by $D^{2}_{\mu} R_v$. 
Eq. (\ref{Polar Decomposition}) has an interesting
physical interpretation that gets a particularly
appealing form (\ref{Polar Decomposition})
in the scattering formalism used here:
The energy derivatives $dT_n/dE$ express the
sensitivity of the conductance to a potential
variation $\Delta U$ on the MC ($d/dE = -\partial/\partial(eU)$).
If this sensitivity is high, the MC is
a fast measuring device to determine
the position of the electron on the DD.
We get a large $R_v$ and therefore a fast
decoherence rate due to the applied voltage $|V|$.
The decoherence rate is thus coupled to the
speed of the measurement process.

The measurement is described by a measurement
time \cite{Aleiner1,Averin1} $\tau_m = 4 S_{II} / (\Delta I)^2$ which
is needed for a signal to noise ratio of 1.
Here $S_{II}$ denotes the low frequency
shot noise spectrum and $\Delta I = I_1-I_2$
is the difference of current flowing through the
MC depending on the state of the two-level system.
This difference is evaluated by use of the
Landauer formula
\begin{equation}
\Delta I = \Delta G |V| = \frac{e^2}{2\pi}
|V| \sum \frac{dT_n}{dE} (e\Delta U)
\end{equation}
where $\Delta G$ is the change of conductance
between the two states of the double dot and
$\Delta U = eC_{\mu}/D(C_i-C_{\mu})$ the potential change on the MC.
The shot noise is as usual
$S_{II} = e|V|(e^2/2\pi)
\sum R_n T_n$. Using
weak coupling $C_1,C_2 \ll C_i$ one gets for
the inverse measurement time
\begin{equation}
\tau_m^{-1} = \Gamma_m = 2\pi \left(\frac{C_{\mu}}{C_i}\right)^2
R_m e|V|
\end{equation}
with the dimensionless constant
\begin{equation}
R_m = \frac{1}{4} 
\frac{\left(\sum \frac{dT_n}{dE}\right)^2}
{\left(\sum \frac{d\phi_n}{dE}\right)^2
 \left(\sum R_n T_n \right)}.
\end{equation}
It is not difficult to show that $R_m \le R_v$ \cite{Pilgram1}
which leads to an important inequality between
measurement rate and decoherence rate $\Gamma_m \le \Gamma_{dec}$.
The measurement is always slower than the
decoherence, the decay of the off-diagonal
elements of the reduced density matrix of the two-level
system.

Which conditions are needed
to get the equality $\tau_m = \Gamma_{dec}^{-1}$?
The tunneling between the two double dots
during the measurement must be negligible,
$\Delta \simeq 0$, and
the temperature must be much smaller than
the applied voltage $kT \ll e|V|$.
More interesting are three constraints
imposed by the scattering matrix
(let the MC be defined by an equilibrium electrostatic potential 
V(x,y,z); let the scattering states be extended
in the z-direction and confined in the xy-plane):

$\bullet$
In order to have $d\phi_{B,n} / dE=0$
in Eq. (\ref{Polar Decomposition})
the scattering Hamiltonian must obey time-reversal
symmetry.

$\bullet$
Furthermore, the 
derivatives $d\phi_{A,n} / dE$ have to vanish.
This can be the case accidentally but is
always fulfilled 
for symmetric detectors 
that obey an inversion symmetry
$V(x,y,z)=V(x,y,-z)$. This condition is well known, 
see for instance \cite{Averin1}.
A quantum-limited measurement can only be
reached by a spacially symmetric detector.
Otherwise part of the information about the state
of the DD is transferred to the phase of the
scattered electrons. This phase does not influence a
conductance measurement. 

$\bullet$
In the multichannel case $N > 1$ another condition
is needed! The equality $R_m = R_v$ then implies
that
\begin{equation}
\label{Multichannel Condition}
\frac{dT_n / dE}{R_n T_n} = C(E).
\end{equation}
The function $C(E) > 0$ does not depend on the
index $n$!
This restriction is of statistical origin: The total
conductance of the detector is a sum of one channel
conductances that have independent uncertainties.
Under condition (\ref{Multichannel Condition})
the statistical uncertainty of their sum 
is minimized.
Eq. (\ref{Multichannel
Condition}) can be interpreted as differential
equations for the transmission probabilities $T_n$.
Their solutions are all of the form
$T_n = (1+e^{-\left(F(E)-F(E_n)\right)})^{-1}$
with $dF/dE = C$ (The function $F$ is therefore monotonously
increasing).
The only difference allowed between
the different probabilities $T_n$
is the offset energy $E_n$.

The simplest case is that of a tunnel contact:
In this case the probabilities $T_n$ 
are dominated by the action 
in the forbidden region and Eqs. (\ref{Multichannel Condition})
are independent of the channel number. 
Thus a tunnel barrier is an efficient detector but has the drawback 
that its measurement time is long.  

Detectors with shorter measurement times can be achieved 
in structures with higher transparencies. 
For such structures the condition Eq. (\ref{Multichannel Condition})
is now important. 
Transmission probabilities of the type (\ref{Multichannel Condition})
occur automatically if
the scattering
problem is separable due to a potential of shape
\begin{equation}
\label{Transmission Condition}
$V(x,y,z) = Z(z) + W(x,y)$.
\end{equation}
This occurs e.g. for the case
$F = 2\pi E/ \omega_z$ with a symmetric 
harmonic scattering potential $Z(z) = V_0 - m\omega_z^2 z^2 /2$.

We demonstrate the importance of
our findings with a generic model that violates
condition (\ref{Transmission Condition}). This condition states
that a geometry with a separable potential $V(x,y,z) = Z(z) +
Y(x,y)$ is favorable to obtain an efficient detector in the case
of more than one open channel. It is clear that a chaotic
potential violates condition (\ref{Transmission Condition})
by definition. We expect therefore that chaos reduces drastically
the efficiency $\Gamma_m / \Gamma_v = R_m / R_v$
of a detector with two open channels, but
has little impact on a detector with only one open channel.

\begin{figure}[htb]
\begin{center}
\leavevmode
\psfig{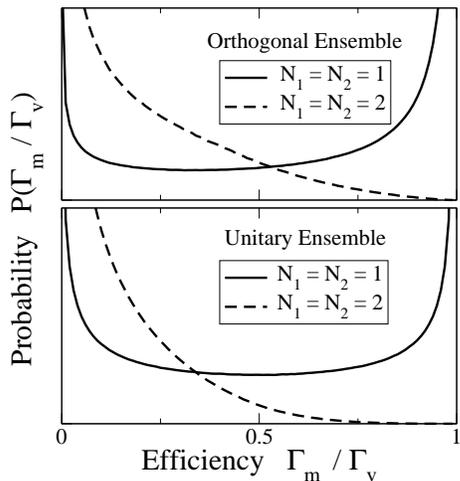}
\caption{Efficiency distribution of an ensemble of chaotic quantum 
cavity detectors: orthogonal ensemble (top panel), unitary ensemble 
(lower panel) for single channel ($N_1 = N_2 = 1$) and 
double channel ($N_1 = N_2 = 2$) point contacts. 
}
\label{Chaos Plot}
\end{center}
\end{figure}

To check this expectation we use a common model of a chaotic cavity
coupled to a left lead with $N_1$ channels and a right lead with $N_2$
channels:
it can be described by a scattering matrix belonging to the circular
ensemble of random matrix theory \cite{Beenakker1}. The distribution
of the density of states matrix elements (\ref{Density Elements}) is
also known \cite{Brouwer1}.
Using these distributions it is straight
forward to obtain the probability distribution of the
measurement efficiency $R_m/R_v$
\begin{equation}
\label{Chaotic Distribution}
\begin{array}{c}
P(R_m/R_v) = \\
\quad\\
\int ds \int dN_E \delta(R_m/R_v - R_m(s,N_E)/R_v(s,N_E)).
\end{array}
\end{equation}
In Eq. (\ref{Chaotic Distribution}) $ds$ is a measure for the circular
en\-sem\-ble of scat\-te\-ring matrices and $dN_E$ a measure for the
symmetrized density of states matrix $N_E = s^{-1/2}ds/dE s^{-1/2}/2\pi i$.
It turns out that the ratio $R_m/R_v$ depends only on the eigenvectors of
$N_E$ and the scattering matrix, but not on the eigenvalues of $N_E$, the
inverse dwell times $\tau_n^{-1}$. The distribution of $R_m/R_v$
is therefore the same in the canonical ($C\ll D$) and grand-canonical
ensemble ($C\gg D$). Their difference is explained in
\cite{Brouwer2}. Fig. \ref{Chaos Plot} shows the
distribution of the measurement efficiency $R_m/R_v$ in the orthogonal 
(time-reversal symmetry) and
unitary ensemble (broken time-reversal symmetry).
The distributions were obtained by numerical integration.
The distribution for $N_1=N_2=1$ in the unitary ensemble can
also be calculated analytically to be 
$P(R_m/R_v)=(R_m/R_v(1-R_m/R_v))^{-1/2}$.
Surprisingly, despite the absence of inversion symmetry 
a chaotic dot with open single channel contacts 
is with high probability an efficient detector!
It is clearly visible that chaos reduces strongly the efficiency of the
measurement device as soon as more than one channel contributes to
the electric transport. The reduction due to a broken time-reversal symmetry
is much less pronounced.

In this work we have analyzed coherent multichannel mesoscopic 
conductors with the aim to find both fast and efficient detectors. 
We find a new statistical condition necessary to carry out
a quantum-limited measurement. This condition relates sensitivities
and shot noises of different conductance channels.
It leads us to a class of detectors 
(defined by separable potentials) that are both fast and efficient. 
We have assumed that a change in the state of the system causes 
only a small change in the potential landscape 
of the detector. Only under this condition 
is it possible to describe the detector response with the help
of small differential changes of the scattering matrix 
and linear screening. We leave it as a future challenging problem 
to develop a theory of phase-coherent, non-linear 
mesoscopic detectors.

This work was supported by the Swiss National Science Foundation.


\begin{thebibliography}{99}
\bibitem{Pilgram1}     The relation $R_m \le R_v$ has the form of
                       a Cauchy-Schwarz inequality.
\end{thebibliography}
\end{document}